\def\half{\frac{1}{2}}
\def\nll{\medskip\noindent} 
\def\hi#1#2{$#1$\kern -2pt-#2} 
\def\hy#1#2{#1-\kern -2pt$#2$} 
\def\dbox#1{\hbox{\vrule 
\vbox{\hrule \vskip #1\hbox{\hskip #1\vbox{\hsize=#1}\hskip #1}\vskip #1 
\hrule}\vrule}}  
\def\qed{\begin{flushright}~{\dbox{0.05true in}}\end{flushright}} 
\def\heading#1{\hfil\linebreak\noindent{\bf #1}~~}
\def\third{{\scriptstyle {1\over 3}}}
\documentclass{iopart}
\usepackage{graphicx}
\begin{document}
\hspace*{2.5 in}CUQM-121, HEPHY-PUB 840/07\\  %


\vspace*{0.4 in}

\title{Ultrarelativistic $N$-boson systems}
\author{Richard L. Hall}
\address{Department of Mathematics and Statistics, Concordia University,
1455 de Maisonneuve Boulevard West, Montr\'eal,
Qu\'ebec, Canada H3G 1M8}
\author{Wolfgang Lucha}
\address{Institute for High Energy Physics, Austrian Academy of Sciences,
Nikolsdorfergasse 18, A-1050 Vienna, Austria}
\eads{\mailto {rhall@mathstat.concordia.ca}, \mailto {wolfgang.lucha@oeaw.ac.at}}
\begin{abstract}General analytic energy bounds are derived for $N$-boson systems governed by
ultrarelativistic Hamiltonians of the~form 
$$H=\sum_{i=1}^N\|{\bf p}_i\|+\sum_{1=i<j}^NV(r_{ij}),$$
 where $V(r)$ is a static attractive pair potential.
It is proved that a translation-invariant model Hamiltonian $H_c$ provides a lower bound to $H$ for all $N \ge 2.$ This result was conjectured in an earlier paper but proved only for $N = 2,3,4.$ As an example, the energy in the case of the linear potential $V(r) = r$ is determined with error less than $0.55\%$ for all $N\ge2.$
\end{abstract}
\vskip0.2in
\maketitle
\section{Introduction}
We consider first the semirelativistic $N$-body Hamiltonian $H$ given by
$$H=\sum_{i=1}^N\sqrt{\|{\mathbf p}_i\|^2+m^2}+\sum_{1=i<j}^NV(r_{ij}),\eqno{(1)}$$
and the following model Hamiltonian $H_c$ 
$$H_c= \sum_{1=i<j}^N\left[{\gamma}^{-1}\sqrt{\gamma\|{\mathbf p}_i-{\mathbf p}_j\|^2+ (mN)^2}~+~ V(r_{ij}) \right],\eqno{(2)}$$
where $\gamma = {N\choose 2} = \half N(N-1).$
If $\Psi(\rho_2,\rho_3,\dots,\rho_N)$ is the lowest boson eigenstate of $H$ expressed in terms of Jacobi relative coordinates, then it was proved in Ref.~\cite{HallLucha2007} that the model facilitates a `reduction' $\langle H_c\rangle = \langle{\mathcal H}\rangle$ to the expectation of a one-body Hamiltonian ${\mathcal H}$ given by
$${\cal H} = N\sqrt{\lambda p^2+ m^2}~~+~\gamma V(r),\quad \lambda ={{2(N-1)}\over{N}}. \eqno{(3)}$$
The question remains as to the relation between $H$ and the model $H_c.$ It is known from earlier work (discussed in~\cite{HallLucha2007}) that the {\bf lower bound conjecture}
$$\langle H\rangle \geq \langle H_c\rangle\eqno{(4)}$$
is true for the following cases: for the armonic oscillator $V(r) = vr^2,$ for all attractive $V(r)$ in the nonrelativistic large-$m$ limit, and for static gravity $V(r) = -v/r$.  This list was augmented in Ref.~\cite{HallLucha2007} by the following cases: in general for $N = 3$, and, if $m = 0,$ for $N = 4.$  The purpose of the present article is to extend this list to include the ultrarelativistic cases $m = 0$ for all $N\ge 2$ and arbitrary attractive $V(r).$
\section{The general lower bound for $m = 0.$}

It was shown in Ref.~\cite{HallLucha2007} that the non-negativity of the expectation $\langle\delta(m,N)\rangle$ is sufficient to establish the validity of the conjecture (4), where
$$
\delta(m, N) = \sum_{i = 1}^{N}\sqrt{\|{\mathbf p}_i\|^2 + m^2}\  -\  {{2}\over{N-1}}\sum_{1=i<j}^{N}
 \sqrt{{{N-1}\over{2N}}\|{\mathbf p}_i -{\mathbf p}_j\|^2+ m^2}.\eqno{(5)}
$$
Thus for the new cases we are now able to treat we must consider $\langle\delta(0,N)\rangle$. By using the necessary boson permutation symmetry of $\Psi$, the expectation value we need to study is reduced to
$$
\langle\delta(0, N)\rangle = N\left\langle\|{\mathbf p}_1\|\  -\ \sqrt{{{N-1}\over{2N}}}\|{\mathbf p}_1 -{\mathbf p}_2\|\right\rangle.\eqno{(6)}
$$
\clearpage 
The principal result of this paper, the lower bound for $m = 0$ and all $N\ge2$, is an immediate consequence of the following:

\nll{\bf Theorem~1}~~~$\langle\delta(0, N)\rangle= 0$.

\nll{\bf Proof of Theorem~1}

\nll Without loss of generality we adopt in momentum space a coordinate origin such 
that $\sum_{i=1}^N{\mathbf p}_i := {\mathbf p}  = {\mathbf 0}.$
We define the mean lengths
$$\langle ||{\mathbf p}_1||\rangle := k \quad {\rm and} \quad \langle ||{\mathbf p}_1-{\mathbf p}_2||\rangle := d.\eqno{(7)}$$
We wish to make a correspondence between mean lengths such as $k$ and $d$ and the sides of triangles that can be constructed with these lengths.  We consider the triangle formed by the three vectors
$\{{\mathbf p}_1, {\mathbf p}_2,~ {\mathbf p}_1-{\mathbf p}_2\}.$ We suppose that the corresponding angles in this triangle are $\{\phi_{12}, \theta_1,\theta_2\}$ (the same notation is used for other similar triples). We now consider projections of one side on a unit vector along an adjacent side and define the mean angles $\phi$ and $\theta$ by the relations
$$\langle \|{\mathbf p}_1\|\cos(\phi_{12})\rangle := \langle \|{\mathbf p}_1\|\rangle\cos(\phi)$$
and
$$\langle\|{\mathbf p}_1-{\mathbf p}_2\|\cos(\theta_1)\rangle := \langle\|{\mathbf p}_1-{\mathbf p}_2\|\rangle\cos(\theta).$$
Thus, on the average, this triangle is isosceles with one angle $\phi$ and the other two angles $\theta.$  Since ${\mathbf p} = 0,$ we have $\langle {\mathbf p}_1\cdot {\mathbf p}\rangle = 0.$  Hence
$$||{\mathbf p}_1||^2 + \sum_{i = 2}^N||{\mathbf p}_1||||{\mathbf p}_i||\cos(\phi_{1i})= 0.$$
Thus, by dividing by $||{\mathbf p}_1||$ and using boson symmetry, we find
$$\langle\left(||{\mathbf p}_1|| + (N-1)||{\mathbf p}_2||\cos(\phi_{12})\right)\rangle = \langle||{\mathbf p}_1||\left(1 + (N-1)\cos(\phi_{12})\right)\rangle= 0. $$
We therefore conclude that $k(1+(N-1)\cos(\phi)) = 0,$ that is to say
$$\cos(\phi) = -\frac{1}{N-1}.$$
We now consider again the triangle formed by the three vectors $\{{\mathbf p}_1, {\mathbf p}_2,~{\mathbf p}_1-{\mathbf p}_2\}.$  We have immediately from the dot product ${\mathbf p}_1\cdot ({\mathbf p}_1-{\mathbf p_2})$
$$\|{\mathbf p}_1\| \|{\mathbf p}_1-{\mathbf p_2}\|\cos(\theta_1) = \|{\mathbf p}_1\|(\|{\mathbf p}_1\| - \|{\mathbf p}_2\|\cos(\phi_{12})).$$
By dividing by $\|{\mathbf p}_1\|$ and taking means we obtain
$$d\cos(\theta)  = k(1-\cos(\phi)).$$
But $\theta = (\pi/2 - \phi/2)$ and $\cos(\phi) = -1/(N-1).$  Hence we conclude
$$\frac{k}{d} = \left(\frac{N-1}{2N}\right)^{\half}.$$
This equality establishes Theorem~1.\qed
\section{The linear potential}
We apply the new bound to the case of the linear potential $V(r) = r.$  The weaker $N/2$ lower bound (discussed in Ref.~\cite{HallLucha2007}) is always available, but, up to now, we knew no way of obtaining tight bounds for this problem. For a comparison upper bound, we use a Gaussian trial function $\Phi$ and the original Hamiltonian $H$ to obtain a scale-optimized variational upper bound $E \leq E_g = (\Phi, H\Phi).$
As we showed in Ref.~\cite{HallLucha2007}, for the linear potential $V(r) = r$ in three spatial dimensions, the conjecture (now proven) implies that the $N$-body bounds are given for $N \ge 2$ by
$$  N\left(\frac{(N-1)^3}{2N}\right)^{1\over 4}e = E_c^L \leq E \leq E_g^U = 4N\left(\frac{(N-1)^3}{2N\pi^2}\right)^{\frac{1}{4}},\eqno{(8)}$$
where $e \approx 2.2322 $ is the bottom of the spectrum~\cite{Boukraa89} of the one-body problem $h = \|{\mathbf p}\| + r$. From (8) we see that the ratio $R = E_g/E_c = 4/(\pi^{\half}e) \approx 1.011.$
The energy of the ultrarelativistic many-body system with linear pair potentials is therefore determined by these inequalities with error less than 0.55\% for all $N\ge 2.$  Earlier we were able to obtain such close bounds for all $N$  only for the harmonic oscillator~\cite{Hall04}.
\section{Conclusion}
We have enlarged the number of semirelativistic problems that satisfy the lower-bound conjecture $\langle H\rangle \geq \langle H_c\rangle$ to include all problems with $m = 0$ and $N\ge 2.$  An extension of the geometric reasoning used in Ref.~\cite{HallLucha2007} from pyramids to more general simplices would perhaps have provided an alternative proof.  However, the more algebraic approach adopted here, relying in the end on mean angles in a triangle, seemed to provide a more independent and robust approach.

 \section*{Acknowledgement}
One of us (RLH) gratefully acknowledges both partial
financial support of his research under Grant No.~GP3438 from~the Natural
Sciences and Engineering Research Council of Canada and hospitality of the
Institute for High Energy Physics of the Austrian Academy of Sciences in
Vienna.
\medskip

\end{document}